\documentclass[twocolumn,prd,nofootinbib,aps,prd,floats,floatfix,superscriptaddress,amsmath,amssymb,longbibliography, twocolumn, secnumarabic]{revtex4-2} %
\usepackage[final]{graphicx}
\usepackage{hyperref}
\usepackage{amsmath}
\usepackage{bbm}
\usepackage{amsfonts}
\usepackage{amssymb}
\usepackage{latexsym}
\usepackage{graphicx}
\usepackage[english]{babel}
\usepackage{multirow}
\usepackage{float}
\usepackage{url}
\usepackage{slashed}
\usepackage{xcolor}
\usepackage[utf8]{inputenc}
\usepackage{verbatim}
\usepackage{stackengine}
\usepackage{cancel}
\usepackage{accents}
\usepackage{array}
\usepackage{physics}
\usepackage{bm}
\begin{document}
\rightline{FERMILAB-PUB-26-0092-T}
\title{Electromagnetic Radiation from Cosmic-Ray Scatterings on Relic Neutrinos}
\author{Gonzalo Herrera}
\email{gonzaloh@mit.edu}
\affiliation{Department of Physics and Kavli Institute for Astrophysics and Space Research, Massachusetts Institute of Technology, Cambridge, MA 02139, USA}
\affiliation{Harvard University, Department of Physics and Laboratory for Particle Physics and Cosmology, Cambridge, MA 02138, USA}
\author{Abraham Loeb}
\email{aloeb@cfa.harvard.edu}
\affiliation{Institute for Theory and Computation, Harvard-Smithsonian Center for Astrophysics, 60
Garden Street, Cambridge, MA 02138, USA}
\begin{abstract}
Cosmic-ray scatterings on the cosmic neutrino background induce a flux of gamma-rays and X-rays from boosted meson decays and charged lepton processes. Here we present the first estimate of this flux and its cumulative cosmological contribution. Confronting expectations with Fermi-LAT diffuse gamma-ray data, we find a limit on the cosmic neutrino background overdensity at the level of $\eta \lesssim 2 \times 10^{4}$ for a lightest neutrino mass of $m_{\nu}\gtrsim0.1$~eV, orders of magnitude stronger than current direct laboratory probes, and comparable to constraints on the cosmic neutrino background obtained with IceCube. We further show that X-ray synchrotron emission from cascade electron-positron pairs in intergalactic magnetic fields provides a complementary, albeit weaker, constraint. We discuss how anisotropic signatures and future gamma-ray data from CTA could further improve bounds on the relic neutrino overdensity, approaching in sensitivity the $\Lambda$CDM expected density.
\end{abstract}
\maketitle

\section{Introduction}

Detecting relic neutrinos from the early Universe remains one of the most important open challenges in astroparticle physics. The cosmic neutrino background (C$\nu$B), a robust prediction of the standard cosmological model, decoupled when the Universe was approximately one second old, at temperatures of order MeV. This occurs well before the main epoch of Big Bang Nucleosynthesis and several orders of magnitude earlier than photon decoupling at recombination. To date, its existence has only been inferred indirectly through its contribution to the radiation energy density during BBN and recombination~\cite{Planck:2018vyg}. This provides indirect gravitational evidence for relic neutrinos at two specific redshifts while leaving their particle properties and subsequent evolution in the Universe experimentally untested.

Direct detection of the C$\nu$B is notoriously difficult because relic neutrinos today are extremely cold, with temperature
\begin{equation}
T_\nu = \left(\frac{4}{11}\right)^{1/3} T_\gamma \simeq 0.17\,\mathrm{meV},
\end{equation}
and number density
\begin{equation}
n_\nu \simeq 336\,\mathrm{cm^{-3}}.
\end{equation}
At such energies, Standard Model interaction cross sections are vanishingly small, rendering conventional detection strategies, for instance scatterings off nuclei or electrons at underground detectors, very ineffective.

One proposed laboratory strategy is neutrino capture on beta-unstable nuclei~\cite{Weinberg:1962zza}, which benefits from the absence of an energy threshold. Experiments such as KATRIN and PTOLEMY aim to probe this possibility~\cite{KATRIN:2022kkv,PTOLEMY:2019hkd}. However, the achievable sensitivity is limited by the extremely small expected event rates and by intrinsic uncertainties in distinguishing capture electrons from the beta-decay endpoint~\cite{PTOLEMY:2022ldz,Cheipesh:2021fmg,Nussinov:2021zrj}.

Recently, it has been recognized that high-energy cosmic rays propagating through the Universe inevitably scatter off relic neutrinos, directly boosting these to high energies, in the range of neutrino telescopes~\cite{Ciscar-Monsalvatje:2024tvm,DeMarchi:2024zer,Herrera:2024upj,Zhang:2025rqh, herrera2026cosmicneutrinobackgroundreach}. The absence of such a signal in IceCube data already constrains the relic neutrino overdensity on cosmological scales to $\eta \lesssim 10^4$ under reasonable assumptions on the cosmic-ray source evolution~\cite{Herrera:2024upj, herrera2026cosmicneutrinobackgroundreach}.

An essential aspect of this scenario is that cosmic-ray interactions with relic neutrinos enter the deep-inelastic regime at high energies. In this regime, neutrino-proton scattering produces hadronic final states that include neutral and charged pions. The decay of neutral pions generates high-energy gamma rays, implying that any boosted-neutrino signal must be accompanied by an electromagnetic counterpart. This connection follows directly from energy conservation and hadronic cascade physics, and therefore provides an independent and unavoidable probe of relic neutrinos in the Universe that complements the direct neutrino detection channel from \cite{Ciscar-Monsalvatje:2024tvm, DeMarchi:2024zer, Herrera:2024upj, Zhang:2025rqh, herrera2026cosmicneutrinobackgroundreach}.

This motivates a multi-messenger approach for the detection of the cosmic neutrino background, in which neutrino-telescope limits are complemented by gamma-ray observations, predominantly from Fermi-LAT \cite{Fermi-LAT:2014ryh}. In addition to gamma rays, the same hadronic interactions produce relativistic electron-positron pairs through charged pion decay and charged-current lepton production. In the presence of extragalactic magnetic fields, these $e^{\pm}$ pairs emit synchrotron radiation at X-ray energies. The cosmic X-ray background (CXB), measured with high precision by HEAO-1~\cite{Gruber:1999yr} and confirmed at comparable levels by INTEGRAL~\cite{Churazov:2006bk} and Swift-BAT~\cite{Ajello:2008xb}, and most recently
NuSTAR~\cite{Krivonos:2020qvl}, therefore provides a complementary window for constraining relic neutrino fluxes. However, as we will show later, the resulting limits are considerably weaker than those from gamma rays due to the subdominant role of synchrotron cooling in typical extragalactic magnetic fields.

Here we estimate the gamma-ray and X-ray fluxes produced by cosmic-ray scattering with the cosmic neutrino background across redshift and derive new constraints on the relic neutrino abundance using Fermi-LAT and HEAO-1 data. We first obtain a conservative bound based on energy-budget considerations with minimal astrophysical assumptions (Section~\ref{sec:estimate}). We then perform a spectral calculation of the gamma-ray and X-ray signals under different cosmic-ray source evolutions and compositions, deriving improved limits on the cosmic neutrino background (Section~\ref{sec:spectral_limit}). In Section~\ref{sec:comparison}, we compare with existing neutrino telescope constraints, discuss anisotropic source contributions, and outline prospects for future multi-messenger searches for relic neutrinos. We present our conclusions in Section~\ref{sec:conclusions}.

\section{A simple estimate}\label{sec:estimate}

Let us first derive an approximate and conservative limit from energy budget arguments. In terms of the total electromagnetic energy density, we can approximate that all the extragalactic background light (EBL) absorbed very high energy gamma-ray emission cascades into the Fermi MeV-GeV band, then
\begin{align}
u_{\gamma}^{(p\nu)} &\simeq \eta \,\kappa_{\rm EM}\,\langle \tau_{p\nu}\rangle\,u_{\rm UHECR},
\end{align}
where $\eta$ is an overdensity factor we wish to constrain. $\kappa_{\rm EM}\sim 0.3$ is the fraction of energy carried by neutral pions that goes into photons,
$u_{\rm UHECR}$ is the ultra-high-energy cosmic-ray (UHECR) energy density, estimated as~\cite{Waxman:1998yy}
\begin{equation}
u_{\rm UHECR} \sim \dot\varepsilon_{\rm CR}\, t_H\sim10^{44}\,\mathrm{erg\,Mpc^{-3}\,yr^{-1}} \times 10^{10} \mathrm{yr},
\end{equation}
and $\langle \tau_{p\nu}\rangle$ is the effective optical depth, which we define as
\begin{equation}
\left\langle\tau_{p \nu}\right\rangle \simeq \bar{n}_\nu(0)\,\sigma_{p \nu}\, c\, t_H\,,
\end{equation}
where $t_H \sim \int d z\, c /[H(z)(1+z)]$. At redshift $z\simeq0$, the relic neutrino background is expected to have an average cosmological number density of $n_{\nu}(z=0)=336$~cm$^{-3}$, summing over all flavors and neutrinos plus antineutrinos~\cite{Dolgov:1997mb}. For the typical center-of-mass energy of UHECR scattering off relic neutrinos, $\sqrt{s}\sim \sqrt{2 m_{\nu}E_{\rm UHECR}} \sim $~GeV, the cross section is $\sigma_{p\nu} \sim 10^{-37}$~cm$^{2}$~\cite{Formaggio:2012cpf, Casper:2002sd}.

The observed isotropic $\gamma$-ray background (IGRB) measured by Fermi-LAT corresponds to an energy density of order $u_\gamma^{\rm IGRB} \sim 10^{-7}\,\mathrm{eV\,cm^{-3}}$~\cite{Fermi-LAT:2014ryh}. Requiring $u_{\gamma}^{(p\nu)} \lesssim 0.1 \times u_\gamma^{\rm IGRB}$ yields an upper limit on the relic neutrino overdensity $\eta$ of
\begin{align}
\eta \lesssim \frac{0.1 \, u_\gamma^{\rm IGRB}}{\kappa_{\rm EM}\,\langle \tau_{p\nu}\rangle\,u_{\rm UHECR}} \lesssim 10^{7},
\end{align}
which is $\sim 4-5$ orders of magnitude better than KATRIN, and about $2-3$ of magnitude worse than IceCube limits from Ref.~\cite{Herrera:2024upj, herrera2026cosmicneutrinobackgroundreach}. In the next section, we refine this calculation, accounting for the full DIS proton-neutrino cross section, spectral shape of the UHECR flux, UHECR emissivity evolving with redshift, energy losses on the EBL, comparison with the IGRB flux from Fermi-LAT at different energies~\cite{Fermi-LAT:2014ryh}, X-ray synchrotron emission, and directional anisotropic effects.

\section{Cosmological diffuse electromagnetic flux from $p\nu$ collisions}\label{sec:spectral_limit}

The estimate from the previous section can be improved in several ways. We begin by modeling the evolution of the ultra-high-energy cosmic-ray flux with redshift in detail. Only ultra-high-energy cosmic rays are able to yield electromagnetic radiation from scattering off relic neutrinos. Pion production in $p\nu$ scattering requires a center-of-mass energy above the pion mass, $\sqrt{s}\gtrsim m_\pi$, which for relic neutrinos of mass $m_\nu\sim 0.1\,{\rm eV}$ implies UHECR protons with lab-frame energies $E_p \gtrsim m_\pi^2/(2m_\nu)\sim 10^{17}\,{\rm eV}$; hence only the highest-energy cosmic rays contribute to this process.

We assume a pure proton composition, with a local comoving energy injection rate of ultra-high-energy cosmic rays of~\cite{Waxman:1998yy}
\begin{equation}
\dot\varepsilon_{\rm CR}(z=0) \simeq 5\times 10^{44}\,{\rm erg\,Mpc^{-3}\,yr^{-1}}.
\end{equation}
We then assume that the comoving cosmic-ray emissivity evolves with redshift as
\begin{equation}
\dot\varepsilon_{\rm CR}(z) = \dot\varepsilon_{\rm CR}(z=0)\, \xi(z)\,,
\end{equation}
following either the Star Formation Rate (SFR)~\cite{Hopkins:2006bw} or the Quasar Evolution Rate (QSO)~\cite{Hopkins:2006fq}. Both evolution functions take the phenomenological form
\begin{equation}
\xi(z) = \frac{(1+z)^{\alpha_1}}{1+\left(\frac{1+z}{z_*}\right)^{\alpha_2}}\,,
\label{eq:xi_general}
\end{equation}
normalized to unity at $z=0$. The SFR evolution uses $\alpha_1 = 2.7$, $z_* = 2.9$, and $\alpha_2 = 5.6$, while the QSO evolution adopts $\alpha_1 = 5.0$, $z_* = 1+z_p = 3.2$ (with $z_p = 2.2$), and $\alpha_2 = 11.0$. The QSO model peaks more sharply and at slightly higher redshift than the SFR, leading to a larger integrated signal.

The neutrino-proton scattering cross section in the resonant and Deep-Inelastic scattering regimes ($\sqrt{s} \gtrsim 1$~GeV) is well described theoretically and further measured experimentally within $1$--$10\%$ precision~\cite{Formaggio:2012cpf}. The dominant contribution arises from charged-current interactions,
\begin{equation}
\nu_{\ell}+p \rightarrow \ell^{-}+X, \quad \bar{\nu}_{\ell}+p \rightarrow \ell^{+}+X\,,
\end{equation}
where $X$ denotes any set of final hadrons. The charged-current DIS cross section is given by~\cite{Schmitz1997Neutrinophysik,Giunti:2007ry}
\begin{equation}
\begin{aligned}
\sigma^{\mathrm{CC}}_{p\nu}
&= 2 \sigma^{\mathrm{CC}}_0
\left[
\sum_{q=d, s}\langle x\rangle_q^p
+\frac{1}{3} \sum_{\bar{q}=\bar{u}, \bar{c}}\langle x\rangle_{\bar{q}}^p
\right], \\
\sigma^{\mathrm{CC}}_{p\bar{\nu}}
&= 2 \sigma^{\mathrm{CC}}_0
\left[
\sum_{\bar{q}=\bar{d}, \bar{s}}\langle x\rangle_{\bar{q}}^p
+\frac{1}{3} \sum_{q=u, c}\langle x\rangle_q^p
\right],
\end{aligned}
\end{equation}
with
\begin{equation}
\sigma^{\mathrm{CC}}_0=\frac{G_{\mathrm{F}}^2}{2 \pi}\, s\,\left(1+\frac{Q^2}{m_W^2}\right)^{-2}.
\end{equation}
In these expressions, $s$ denotes the center-of-mass energy squared, and $\langle x\rangle_{q, \bar{q}}^p=\int \mathrm{d} x\, x\, f_{q, \bar{q}}^p(x)$ denote the average fractional momenta carried by quarks and antiquarks~\cite{Giunti:2007ry}. We take these from $\texttt{ManeParse}$~\cite{Clark:2016jgm}. The center-of-mass energy depends explicitly on the neutrino mass through $s \simeq 2 m_\nu E_p$, which determines the threshold and normalization of the cross section.

We define an effective energy-weighted cross section that accounts for the UHECR spectral shape as
\begin{equation}
\langle\sigma_{p \nu}^{\mathrm{eff}}\rangle=\frac{\int_{E_{\min}}^{E_{\max}} d E_p\, E_p^{1-\alpha}\, e^{-E_p / E_{\max}} \,\sigma_{p \nu}\!\left(E_p\right)}{\int_{E_{\min}}^{E_{\max}} d E_p\, E_p^{1-\alpha}\, e^{-E_p / E_{\max}}}\,,
\label{eq:sigma_eff}
\end{equation}
where $\alpha$ is model-dependent. For an SFR-dominated evolution, we take $\alpha=2.2$~\cite{Kotera_2011}, and for QSO, we take $\alpha=2.0$~\cite{Murase:2016gly}. The integration is performed from $E_{\min} = 10^{19}$~eV, above which the $p\nu$ cross section contributes appreciably, up to $E_{\max} = 3 \times 10^{21}$~eV.

\subsection{Gamma-ray flux}

With these ingredients, we construct the line-of-sight electromagnetic emissivity from $p\nu$ scatterings. At each redshift, the comoving UHECR energy density is set by the balance between continuous injection and Hubble dilution~\cite{Waxman:1998yy},
\begin{equation}
u_{\rm CR}(z) \;\simeq\; \frac{\dot\varepsilon_{\rm CR}(z)}{H(z)}\,,
\label{eq:uCR}
\end{equation}
where $H(z) = H_0\sqrt{\Omega_m(1+z)^3 + \Omega_\Lambda}$. This expression captures the approximate steady-state density in an expanding Universe. Cosmic rays are injected at rate $\dot\varepsilon_{\rm CR}$ and diluted on a Hubble timescale $H^{-1}(z)$.

The volume emissivity (electromagnetic energy injection rate per unit comoving volume) for each channel is then
\begin{equation}
j_{i}(z) \;\simeq\; \kappa_{i}\,\eta \, n_\nu(0)\,(1+z)^3\,
\langle\sigma_{p\nu}^{\rm eff}\rangle\, c \,u_{\rm CR}(z)\,,
\label{eq:emissivity}
\end{equation}
where $i=\{\pi^{0}, \pi^{\pm}, \ell^{\pm}\}$, the parameter $\eta\equiv n_\nu/\bar n_\nu$ quantifies a possible overdensity of relic neutrinos relative to the standard cosmological expectation, and the factor $(1+z)^3$ accounts for the increasing physical number density of the C$\nu$B with redshift. The energy fractions $\kappa_i$ denote the fraction of incoming cosmic-ray energy channeled into each electromagnetic final state. At center-of-mass energies of a few GeV, we adopt $\kappa_{\pi^0}\simeq 0.2$, $\kappa_{\pi^\pm}\simeq 0.2$, and $\kappa_{\ell^{\pm}} \simeq 0.1$~\cite{Kelner:2006tc,Anchordoqui:2007tn,Gandhi:1995tf,Giunti:2007ry}; the remaining $\sim 50\%$ of the hadronic energy is carried away by neutrinos.

Neutral pions decay promptly into two photons ($\pi^0\to\gamma\gamma$), while charged pions decay into $\mu^\pm$ which subsequently produce $e^\pm$ pairs. Primary electrons from the charged-current lepton channel are also produced. These electrons and positrons radiate via inverse Compton scattering on the CMB and EBL, further feeding the electromagnetic cascade. We include all these components in our calculation. The differential gamma-ray flux observed at energy $E_\gamma$ is obtained from the line-of-sight integral,
\begin{align}
\frac{d\Phi}{dE_\gamma} &=
\frac{c}{4\pi}\int_0^{z_{\max}} dz\,
\frac{1}{H(z)(1+z)}\,
\sum_{i}\,j_i(z)\,
T_{i}(E_\gamma,z)\,
e^{-\tau_{\gamma\gamma}(E_\gamma,z)}.
\label{eq:flux}
\end{align}
We adopt cosmological parameters $H_0=67.4$~km/s/Mpc, $\Omega_{\Lambda}=0.685$, and $\Omega_m=0.315$~\cite{Planck:2018vyg}. The functions $T(E_\gamma,z)$ are cascade redistribution kernels that
map the injected electromagnetic energy into the observed GeV--TeV band
after reprocessing through electromagnetic cascades on the CMB and
EBL. The reprocessed
photon spectrum may acquire a universal shape, independent of whether the
initial electromagnetic energy was injected as photons (from
$\pi^0\to\gamma\gamma$), electrons (from
$\pi^\pm\to\mu\to e^\pm$), or primary leptons (from charged-current
interactions)~\cite{Berezinsky:2016feh,Murase:2012xs}. We use a single cascade kernel for all three electromagnetic channels, parameterized as
\begin{equation}
\begin{aligned}
T(E_{\gamma},z)
&= \mathcal{N}^{-1}(1+z)\,E_{\gamma}^{-\alpha_{\rm cas}} \\
&\quad \times
\exp\!\left[-\!\left(\frac{E_{\gamma,\min}}{E_{\gamma}}\right)^\beta\right]
\exp\!\left[-\!\left(\frac{E_{\gamma}}{E_c(z)}\right)^\beta\right],
\end{aligned}
\end{equation}
with $\alpha_{\rm cas}\simeq 1.9$ \cite{Berezinsky:2016feh}, $\beta \simeq 3$, and
normalization $\mathcal{N}$ enforcing
$\int dE_\gamma\, E_\gamma\, T = 1$, so that all injected
electromagnetic energy is conserved after cascading. Note that
$\alpha_{\rm cas}$ is determined by the cascade multiplication process
and is independent of the UHECR source spectral index. The
high-energy cutoff $E_c(z) \simeq 300/(1+0.5\,z)$~GeV captures the
energy above which pair production on the EBL efficiently reprocesses
photons, with a redshift dependence reflecting the increasing EBL
opacity at earlier times. The low-energy cutoff
$E_{\gamma,\min}\simeq 0.1$~GeV reflects the minimum energy of
inverse-Compton upscattered CMB photons by cascade electrons. The
channel-dependent energy fractions $\kappa_i$ in Eq.~(15) already
account for how much electromagnetic energy each process injects into
the cascade; the spectral redistribution thereafter is universal. Our
kernel parameterization is phenomenological. A full numerical cascade
treatment, e.g.\ with \texttt{CRPropa}~\cite{CRPropa:2022ovg} or
$\gamma$\texttt{-Cascade}~\cite{Blanco:2018bbf}, would provide a more
rigorous spectral prediction. Having said that, the bounds on $\eta$
derived here are primarily sensitive to the total electromagnetic
energy budget, rather than to the detailed spectral redistribution
within the Fermi-LAT band.
The exponential factor $e^{-\tau_{\gamma\gamma}}$ accounts for the absorption of cascade photons on the EBL, which becomes relevant at $\gtrsim 50$--$100$~GeV depending on redshift. We use the EBL model of~\cite{Dom_nguez_2010}.

\begin{figure}[t!]
    \centering
    \includegraphics[width=0.99\linewidth]{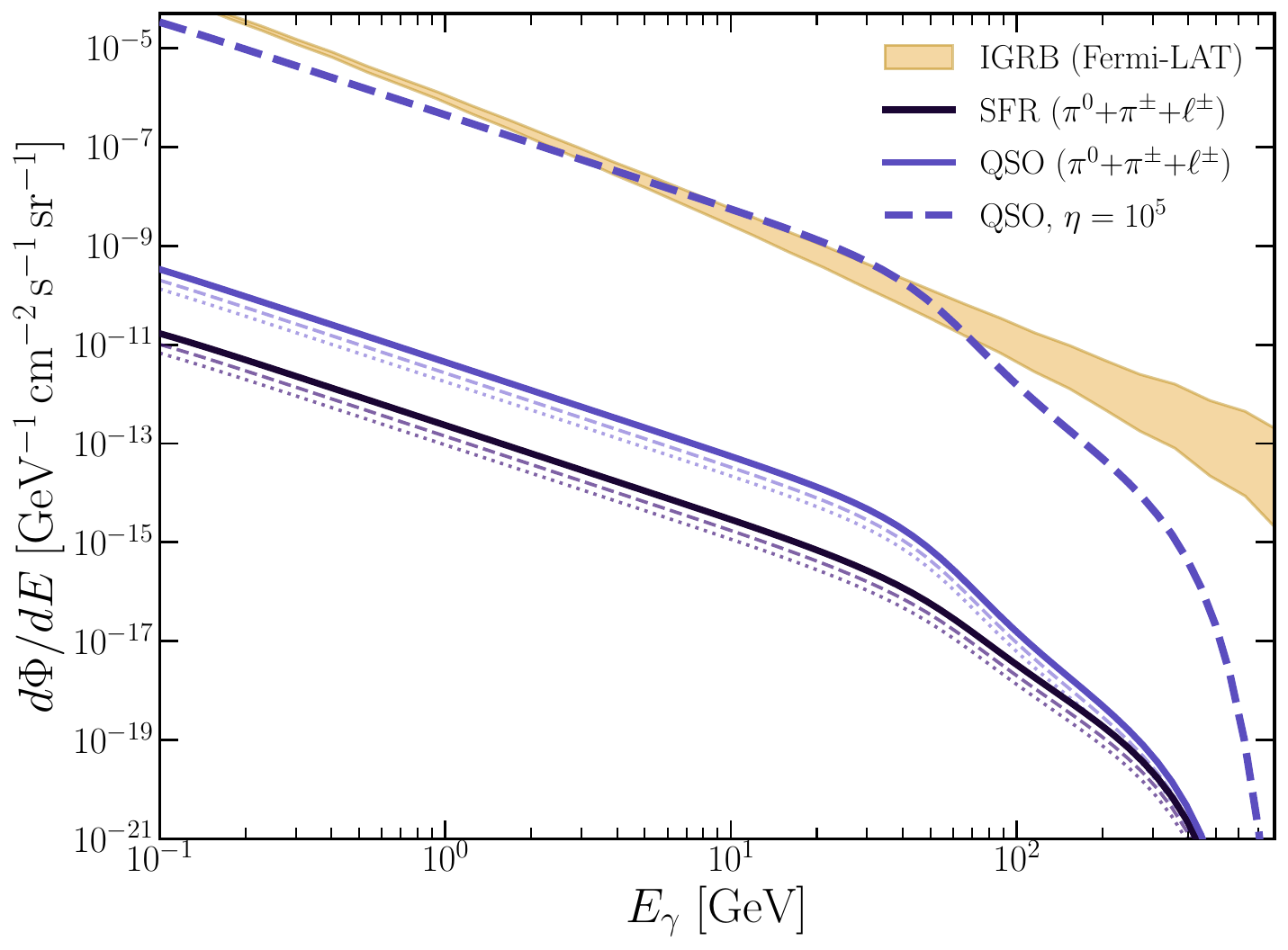}
\caption{Differential gamma-ray flux from cosmic-ray scatterings with the cosmic neutrino background on cosmological scales. Solid lines show the total flux summing all electromagnetic channels: neutral pion decay ($\pi^0\to\gamma\gamma$, dotted), charged pion decay ($\pi^{\pm}\to\mu\to e^{\pm}$, dashed), and charged-current lepton production. Both SFR and QSO models for the cosmic-ray emissivity evolution are shown assuming $\eta=1$ (standard C$\nu$B density), alongside a QSO prediction with local overdensity $\eta=10^5$. The shaded band is the Fermi-LAT isotropic gamma-ray background (IGRB)~\cite{Fermi-LAT:2014ryh}.}
    \label{fig:gamma_spectrum}
\end{figure}

In Figure~\ref{fig:gamma_spectrum}, we show our calculated gamma-ray fluxes from $p\nu$ scatterings, considering both the SFR and QSO evolution models. Solid lines show the total flux from all electromagnetic channels (neutral pions, charged pions, and charged leptons), while the dotted and dashed lines show the individual contributions from the $\pi^0\to\gamma\gamma$ and charged-particle ($\pi^\pm + \ell^\pm$) channels, respectively. We further overlay diffuse IGRB measurements from Fermi-LAT~\cite{Fermi-LAT:2014ryh} for comparison. For the QSO scenario, a C$\nu$B overdensity of $\eta \simeq  10^{4}-10^{5}$ would be required to yield a prediction at the level of the Fermi-LAT measurements, with the strongest sensitivity arising in the 10-50 GeV band in our phenomenological description.

\subsection{X-ray synchrotron flux}\label{sec:xray}

The same hadronic interactions that produce gamma rays also inject relativistic $e^\pm$ pairs through the charged pion ($\pi^\pm \to \mu^\pm \to e^\pm$) and charged-current lepton channels. In the presence of extragalactic magnetic fields, these electrons and positrons cool via both inverse Compton scattering on the CMB and synchrotron radiation. While the inverse Compton component feeds the gamma-ray cascade described above, the synchrotron component produces photons at X-ray energies, providing a complementary observational channel.

The fraction of $e^\pm$ energy radiated as synchrotron versus inverse Compton is determined by the ratio of the magnetic and radiation energy densities~\cite{Blumenthal:1970gc},
\begin{equation}
f_{\rm syn}(B,z) = \frac{u_B}{u_B + u_{\rm rad}}\,,
\label{eq:fsyn}
\end{equation}
where $u_B = B^2/(8\pi)$ is the magnetic energy density and $u_{\rm rad} \simeq u_{\rm CMB}(1+z)^4 \times 1.1$ accounts for the CMB plus a $\sim 10\%$ contribution from EBL starlight at low redshift \cite{Dom_nguez_2010}. At $z=0$, the CMB energy density is $u_{\rm CMB} \simeq 0.26$~eV/cm$^3$.

\begin{figure*}[t!]
    \centering
    \includegraphics[width=0.49\linewidth]{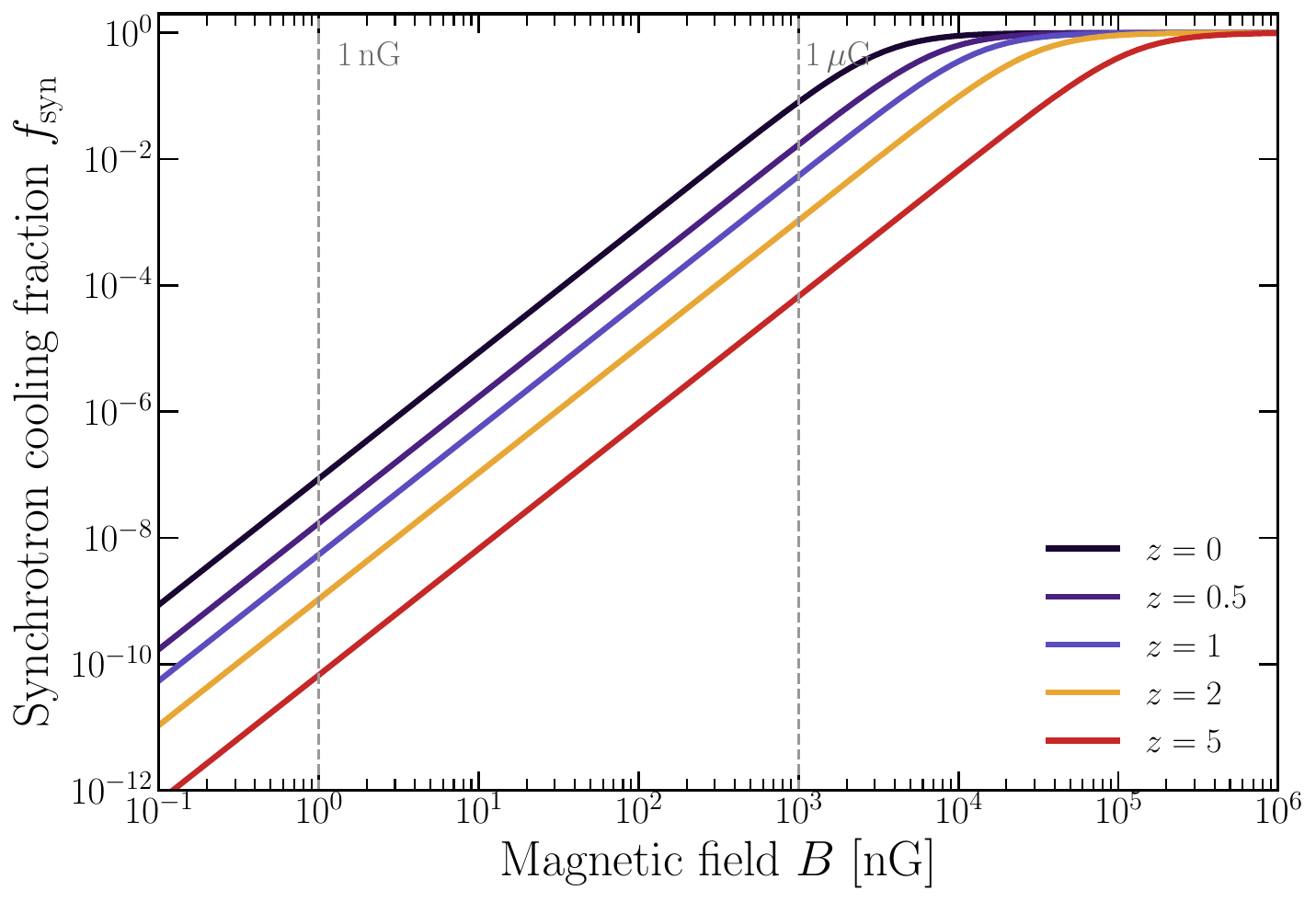}
    \includegraphics[width=0.49\linewidth]{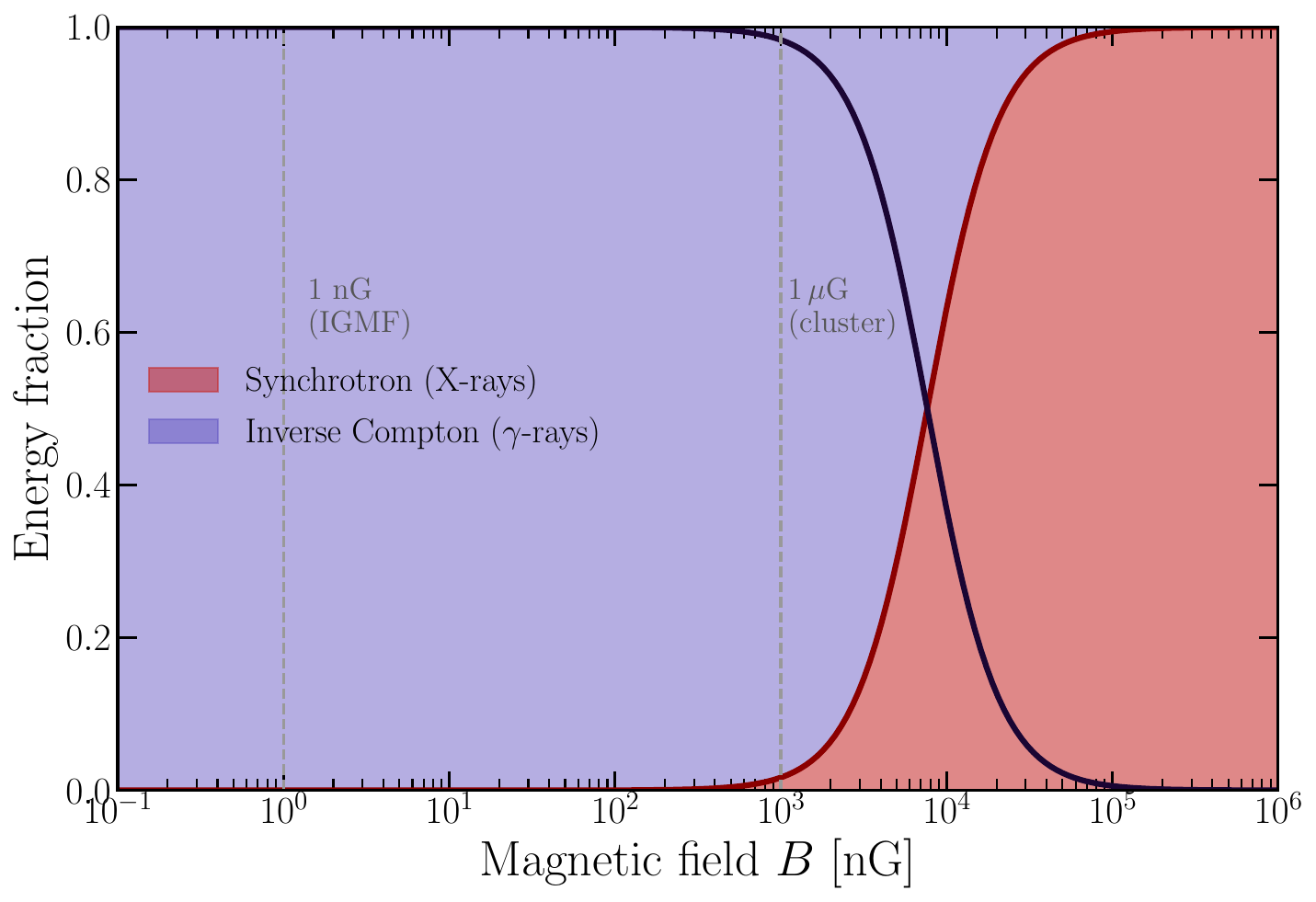}
\caption{%
\textit{Left:} Synchrotron cooling fraction 
$f_{\rm syn} = u_B/(u_B + u_{\rm rad})$ as a function of 
magnetic field strength for several redshifts. At low $z$, 
synchrotron losses begin to dominate above $B \sim 1\,\mu$G; 
at higher redshifts the rising CMB energy density 
($u_{\rm rad} \propto (1+z)^4$) shifts this transition to 
stronger fields.
\textit{Right:} Fractional energy budget of cascade $e^\pm$ 
at $z=0.5$, showing the partition between synchrotron 
(X-ray) and inverse-Compton ($\gamma$-ray) cooling channels. 
For intergalactic fields $B \lesssim 1$~nG virtually all 
energy goes into $\gamma$-rays, while in galaxy-cluster 
cores ($B \sim \mu$G) a significant fraction is redirected 
into X-ray synchrotron emission.}
    \label{fig:sync_cooling}
\end{figure*}

The synchrotron cooling fraction is strongly field-dependent. For intergalactic magnetic fields of $B \sim 1$~nG, representative of the intergalactic medium, one finds $f_{\rm syn} \sim 10^{-7}$; only for $B \gtrsim 10\,\mu$G, characteristic of galaxy cluster cores~\cite{Carilli:2001hj}, does synchrotron dominate over inverse Compton. This hierarchy is important to understand why X-ray constraints on the C$\nu$B are substantially weaker than gamma-ray constraints on cosmological scales, see Fig.~\ref{fig:sync_cooling}. Although strong magnetic fields exist in Galaxies, they posess a small volume filling factor in intergalactic space.

The characteristic synchrotron photon energy produced by an electron of energy $E_e$ in a magnetic field $B$ is~\cite{Blumenthal:1970gc}
\begin{equation}
E_{\rm syn} \;\simeq\; 0.67~{\rm eV}\,
\left(\frac{B}{1~{\rm nG}}\right)\,
\left(\frac{E_e}{100~{\rm TeV}}\right)^2\,
\frac{1}{1+z}\,.
\end{equation}
For cascade electrons with typical energies $E_e \sim 1$-$100$~TeV in intergalactic fields of $B \sim 1\,\mu$G $= 1000$~nG, the synchrotron emission peaks in the keV band, directly in the window of the cosmic X-ray background.

The differential X-ray flux from synchrotron radiation is computed analogously to the gamma-ray flux,
\begin{align}
\frac{d\Phi_X}{dE_X} &=
\frac{c}{4\pi}\int_0^{z_{\max}} dz\,
\frac{1}{H(z)(1+z)}\,
\left(\kappa_{\pi^\pm}+\kappa_{\ell^\pm}\right)\,\eta\, n_\nu(z)
\nonumber\\
&\qquad \times
\langle\sigma_{p\nu}^{\rm eff}\rangle\, c\, u_{\rm CR}(z)\,
f_{\rm syn}(B,z)\,
T_X(E_X,z,B)\,,
\label{eq:xray_flux}
\end{align}
where only the charged channels ($\pi^\pm$ and $\ell^\pm$) contribute, since neutral pion decay produces photons rather than $e^\pm$ pairs. The synchrotron redistribution kernel $T_X(E_X,z,B)$ maps the injected electron energy into the X-ray band. For a single electron of energy $E_e$ in a magnetic field $B$, the radiated synchrotron spectrum is proportional to the standard emission function~\cite{1979rpa..book.....R}
\begin{equation}
F(x) = x \int_x^{\infty} K_{5/3}(\xi)\,d\xi\,,
\label{eq:synchrotron_F}
\end{equation}
where $K_{5/3}$ is the modified Bessel function of the second kind and $x = E_X / E_c(E_e, B, z)$, with the critical synchrotron energy
\begin{equation}
E_c = \frac{3}{2}\,\frac{e\hbar B}{m_e c}\,\gamma_e^2\,\frac{1}{1+z}\,.
\end{equation}
The kernel $T_X$ is constructed by integrating $F(x)$ over a representative cooled electron spectrum $dN/dE_e \propto E_e^{-2}$ with energies $E_e \sim 1$--$300$~TeV, and is normalized such that $\int dE_X\, E_X\, T_X = 1$, consistent with the gamma-ray employed kernels.

\begin{figure}[t!]
    \centering
    \includegraphics[width=0.99\linewidth]{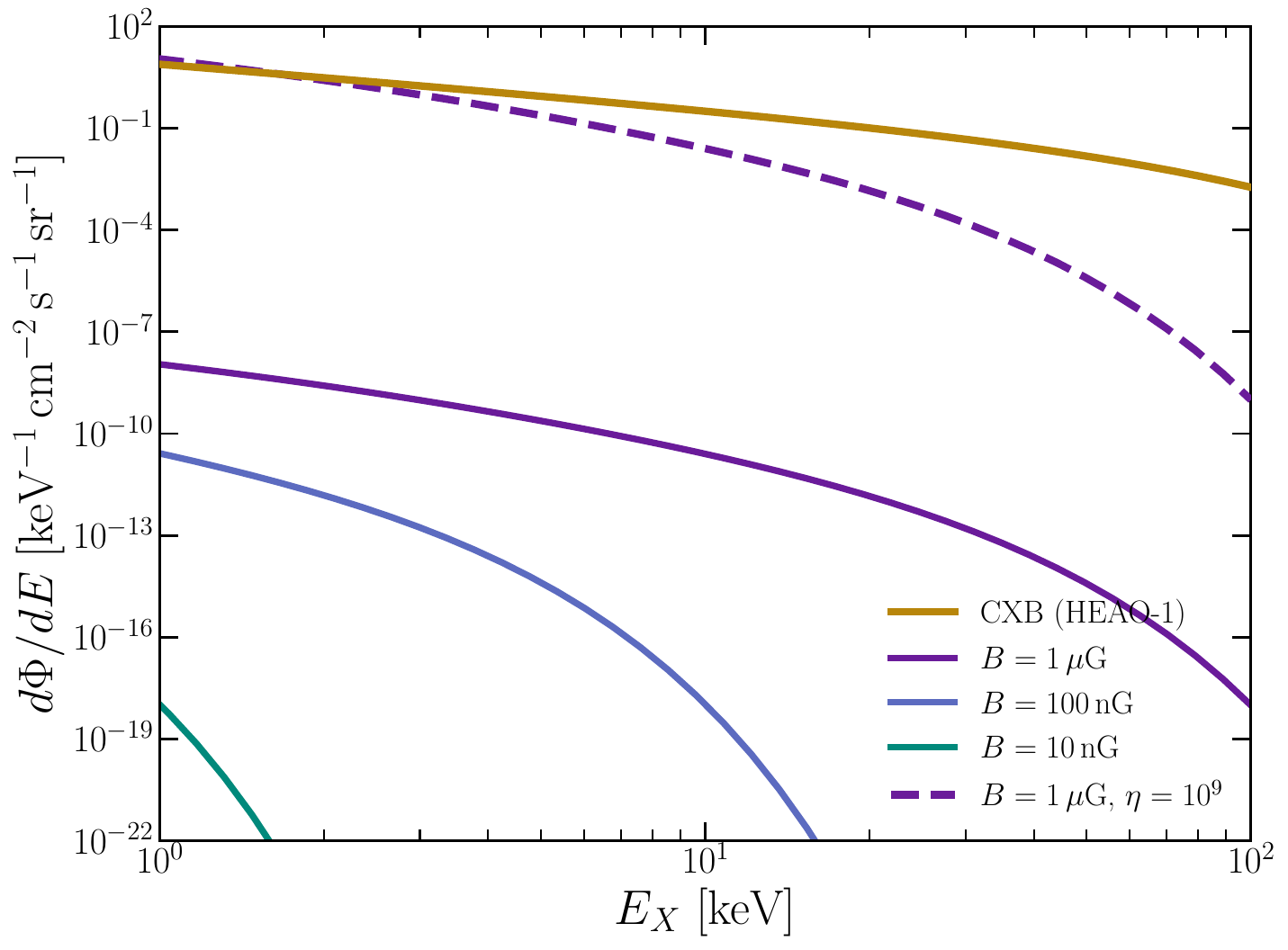}
\caption{Differential X-ray flux from synchrotron radiation of cascade $e^{\pm}$ pairs produced in cosmic-ray scatterings with the C$\nu$B. Solid lines show the QSO evolution model and dashed lines the SFR model, for intergalactic magnetic field strengths $B = 1\,\mu$G, $100\,$nG, and $10\,$nG, all assuming the $\Lambda$CDM-expected C$\nu$B density ($\eta = 1$). A QSO prediction with $B = 1\,\mu$G and local overdensity $\eta = 10^9$ is shown for comparison. The solid gold curve is the cosmic X-ray background (CXB) spectrum measured by HEAO-1~\cite{Gruber:1999yr}. The large gap between predictions and the CXB reflects both the small synchrotron cooling fraction $f_{\rm syn} \ll 1$ in typical fields and the intrinsic brightness of the CXB.}
    \label{fig:xray_spectrum}
\end{figure}

In Figure~\ref{fig:xray_spectrum}, we show the predicted X-ray synchrotron flux for several magnetic field strengths alongside the CXB spectrum measured by HEAO-1~\cite{Gruber:1999yr} in the 3-60~keV band. Even for an  optimistic case of $B = 1\,\mu$G (characteristic of galaxy cluster), the $\eta=1$ prediction lies about 9 orders of magnitude below the CXB, far weaker than the gamma-ray constraints. The physical origin of this hierarchy is twofold. First, the synchrotron cooling fraction is small, and second, the CXB is intrinsically orders of magnitude brighter than the IGRB in terms of photon number flux.

\subsection{Anisotropic signatures}

Considering source-dependent effects can improve the gamma-ray sensitivity further. The CR--C$\nu$B interaction rate depends on both the relic neutrino density and the cosmic-ray flux along each line of sight. So far we have assumed isotropy to derive a limit, but isotropy is not a good description in the nearby Universe.

The density of massive relic neutrinos is enhanced in the potential wells of the Local Group and nearby galaxy clusters, while the cosmic-ray emissivity traces the inhomogeneous distribution of matter along the supergalactic plane. These effects introduce natural anisotropies in the predicted gamma-ray flux. Quantitatively, the relic neutrino density in the direction of galaxy clusters is larger by a factor of at least $\eta_{\nu} \sim 2$~\cite{Ringwald:2004np,LoVerde:2013lta,deSalas:2017wtt}, potentially larger depending on the halo mass and redshift.

Regarding UHECR anisotropies, the arrival distribution is nearly isotropic, with the Pierre Auger Observatory reporting a dipole amplitude of only a few percent~\cite{PierreAuger:2017pzq}. However, this small observed anisotropy does not reflect the intrinsic anisotropy of the source population. Magnetic deflections in Galactic and extragalactic fields isotropize the arrival directions, even when the underlying source emissivity is strongly inhomogeneous. Numerical simulations in which UHECR sources trace the local large-scale structure produce emissivity contrasts of $\mathcal{O}(1)$ between directions along the supergalactic plane and toward nearby voids~\cite{Sigl:2004yk}. In the following, we assume that the cosmic-ray emissivity can produce a factor of $\eta_{\rm CR}\sim 2$ contrast between the brightest and faintest sky directions.

\begin{figure}[t!]
    \centering
    \includegraphics[width=0.99\linewidth]{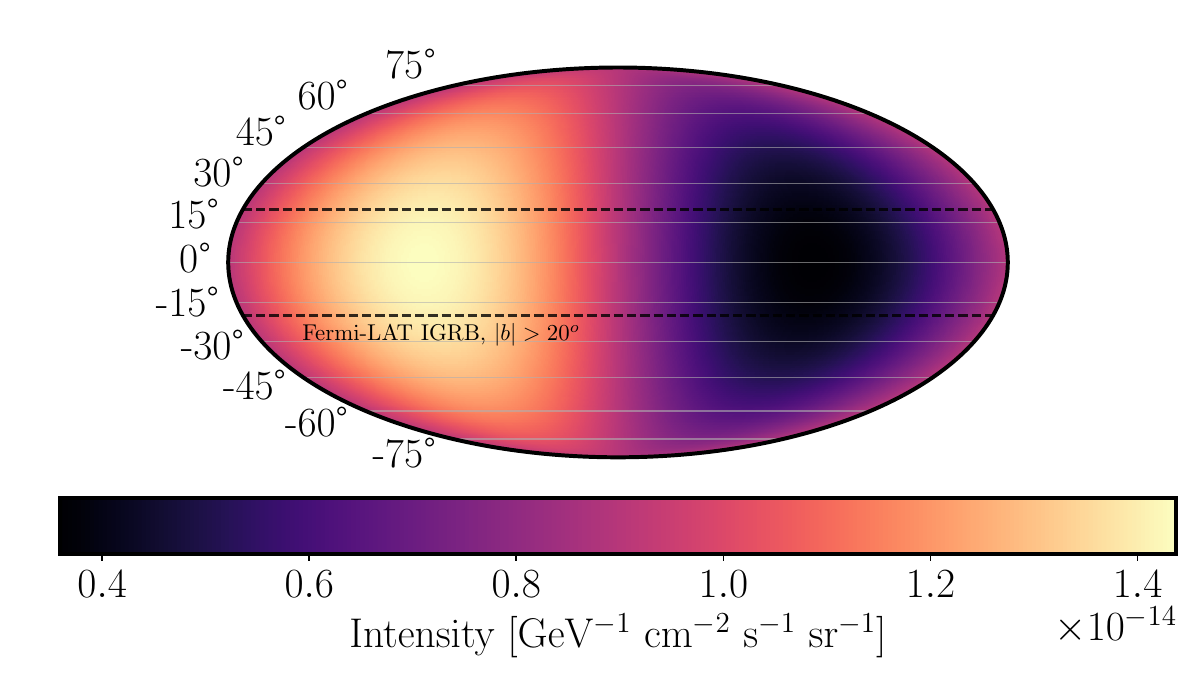}
    \caption{Expected dipole sky anisotropy of the gamma-ray flux induced by cosmic-ray scatterings with the C$\nu$B. The map shows the absolute intensity modulated by the assumed dipole at a reference energy of 10~GeV. The dashed lines indicate $|b|=20^\circ$, delimiting the region used in the Fermi-LAT isotropic background analysis.}
    \label{fig:skymap}
\end{figure}

We describe the resulting large-scale anisotropy using a dipole modulation of the form~\cite{PierreAuger:2017pzq, Fermi-LAT:2012pez}
\begin{equation}
I_{\gamma}(\hat{n}) \;=\; I_{\rm iso}\,\bigl[1 + \delta_{\gamma}\,
(\hat{n}\!\cdot\!\hat{d})\bigr]\,,
\end{equation}
where $I_{\rm iso}$ is the isotropic baseline intensity, $\hat{d}$ is the direction of maximal enhancement, and $\delta_{\gamma}$ is the dipole amplitude. Since the gamma-ray signal scales as $I_\gamma(\hat{n}) \propto n_\nu(\hat{n})\, j_{\mathrm{CR}}(\hat{n})$, the maximum contrast between directions of smallest and largest enhancement is
\begin{equation}
\frac{I_{\rm max}}{I_{\rm min}}
= \frac{1+\delta_{\gamma}}{1-\delta_{\gamma}}
= \eta_{\nu}\,\eta_{\rm CR}\,,
\end{equation}
giving
\begin{equation}
\delta_{\gamma} \;=\; \frac{\eta_{\nu}\,\eta_{\rm CR} - 1}{\eta_{\nu}\,\eta_{\rm CR} + 1}\,.
\end{equation}
For the fiducial example $\eta_{\nu}\,\eta_{\rm CR}=4$, this yields $\delta_{\gamma}\simeq 0.6$, corresponding to an $\mathcal{O}(1)$ dipole anisotropy. To visualize this, we apply the dipole modulation to the computed isotropic CR--C$\nu$B gamma-ray intensity at a reference energy of 10~GeV and generate an all-sky map in Galactic coordinates, shown in Figure~\ref{fig:skymap}. The region used in the Fermi-LAT IGRB analysis (defined by $|b|>20^{\circ}$) is indicated by dashed lines.

Since the Fermi-LAT IGRB analysis excludes the Galactic plane, the effective constraint comes from comparing the $p\nu$ signal against the IGRB in the off-plane sky. The anisotropy implies that a directional analysis can improve the bound on $\eta$ by a factor of $\sim \eta_\nu \eta_{\rm CR} \simeq 4$ relative to the isotropic diffuse limit.

\section{Results and comparison with complementary probes}\label{sec:comparison}

We now present the constraints on the C$\nu$B overdensity $\eta$ obtained from our spectral analysis and compare them with existing bounds from complementary probes.

\subsection{Gamma-ray constraints}

To derive the upper limit on $\eta$ from gamma-ray observations, we require that the predicted $p\nu$ flux does not exceed the 100\% of the Fermi-LAT IGRB at any energy in the measured range~\cite{Fermi-LAT:2014ryh}. This criterion is applied at each energy bin, and the most constraining bin determines $\eta_{\rm max}$. The limit depends on the neutrino mass through the energy-weighted cross section $\langle \sigma_{p\nu}^{\rm eff}\rangle$, which scales linearly with $m_\nu$ in the DIS regime.

Figure~\ref{fig:eta_limits} shows the resulting upper limits on $\eta$ as a function of the lightest neutrino mass $m_\nu$, converting to the heaviest mass state using the normal hierarchy squared mass differences $\Delta m^2_{31} = 2.453 \times 10^{-3}$~eV$^2$~\cite{Esteban:2020cvm}. For the QSO evolution model and $m_\nu = 0.1$~eV, the diffuse Fermi-LAT data yields
\begin{equation}
\eta \lesssim 2.2 \times 10^4 \quad \text{(QSO, Fermi diffuse)}\,,
\end{equation}
while the SFR evolution gives a weaker bound of $\eta \lesssim 10^6$ due to its less peaked source evolution. Including the anisotropy enhancement factor of $\sim 4$ from the directional analysis described in Section~\ref{sec:spectral_limit} strengthens these bounds by a corresponding factor.

\subsection{X-ray constraints}

As discussed in Section~\ref{sec:xray}, the X-ray synchrotron constraints are substantially weaker than the gamma-ray bounds. For a magnetic field $B = 1\,\mu$G, the CXB data constrains $\eta \lesssim 10^{9}$, while for $B = 10$~nG the limit weakens to $\eta \lesssim 10^{13}$. These values are indicated by text annotations in Figure~\ref{fig:eta_limits}. The X-ray channel is thus not competitive with gamma-ray observations for constraining the C$\nu$B overdensity, but it provides a useful consistency check. Any putative detection in gamma rays would predict a specific, testable X-ray signature whose amplitude depends on the ambient magnetic field. In principle, future X-ray missions with improved sensitivity to the diffuse background, combined with observations toward galaxy clusters where $B \sim \mu$G fields are present, could improve these constraints.
\subsection{Comparison with other probes}

Our gamma-ray bounds can be compared with several complementary constraints on the C$\nu$B overdensity. First, the total neutrino energy density today cannot exceed the critical density of the Universe. Writing the relic neutrino number density as $n_\nu = \eta\,\bar n_\nu$, where $\bar n_\nu \simeq 336~\mathrm{cm^{-3}}$ is the standard cosmological average (summed over all mass eigenstates), the requirement
\begin{equation}
\rho_\nu = \eta\,\bar n_\nu\,\langle E_\nu \rangle \le \rho_c,
\qquad 
\rho_c = 3 H_0^2 M_P^2,
\end{equation}
implies
\begin{equation}
\eta_{\max} = 
\frac{\rho_c}{\bar n_\nu\,\langle E_\nu \rangle}.
\end{equation}

Today, relic neutrinos with $m_i \gtrsim T_{\nu,0}$ are non-relativistic, so 
$\langle E_\nu \rangle \simeq \frac{1}{3}\sum_i m_i$, and the bound becomes
\begin{equation}
\eta_{\max} \simeq 
\frac{\rho_c}{(\bar n_\nu/3)\sum_i m_i}
\;\propto\; \frac{1}{\sum_i m_i},
\end{equation}
leading to a mass-dependent upper limit that differs slightly between the normal and inverted hierarchies.

In the opposite limit $m_i \ll T_{\nu,0}$, where relic neutrinos remain relativistic today and 
$\langle E_\nu\rangle \simeq 3.15\,T_{\nu,0}$, the bound saturates at
\begin{equation}
\eta_{\max} \simeq 
\frac{\rho_c}{\bar n_\nu\,3.15\,T_{\nu,0}}
\sim 2 \times 10^4.
\end{equation}

\begin{figure}[t!]
    \centering
    \includegraphics[width=0.99\linewidth]{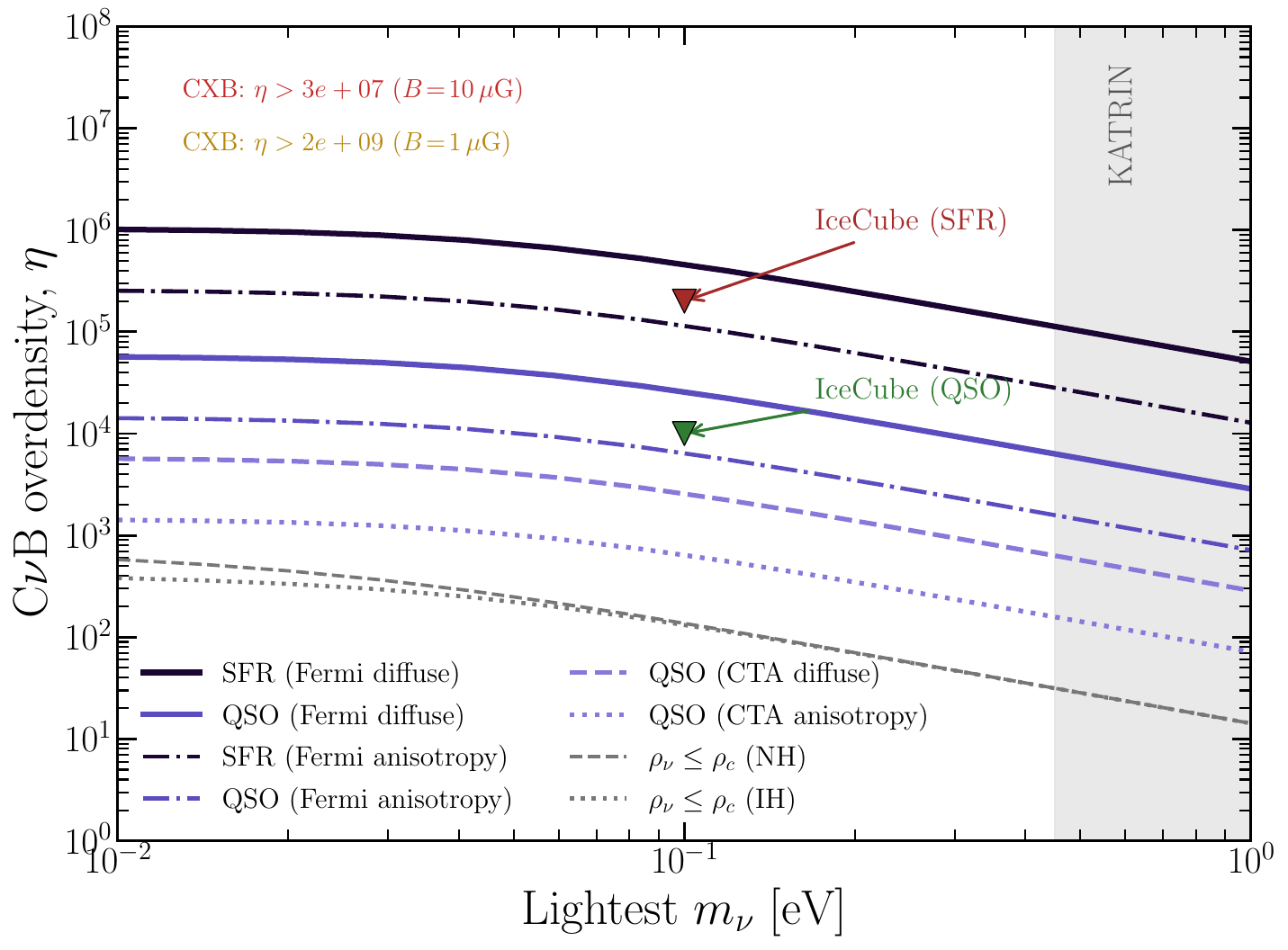}
    \caption{Upper limits on the C$\nu$B overdensity $\eta$ compared to the $\Lambda$CDM expectation, as a function of the lightest neutrino mass. The solid lines are obtained from $p\nu$ scatterings not inducing a gamma-ray flux that overshoots the diffuse IGRB gamma-ray measurements at Fermi-LAT~\cite{Fermi-LAT:2014ryh}, or projected sensitivities at CTA~\cite{CTAConsortium:2010umy}, while the dashed lines are obtained from not observing a large directional anisotropy in Fermi-LAT data, or future CTA data. The black lines correspond to a cosmic-ray evolution following the star forming rate from~\cite{Hopkins:2006bw}, and the purple lines correspond to the quasar evolution function from~\cite{Hopkins:2006fq}.}
    \label{fig:eta_limits}
\end{figure}

In Fig.~\ref{fig:eta_limits}, we show the resulting mass-dependent cosmological bound for both the normal and inverted hierarchies. 

Neutrinos can also cluster gravitationally, independently from Pauli exclusion principle considerations. For currently allowed neutrino masses, gravitational clustering in galaxy-scale halos enhances the local relic neutrino density by at most a factor of $\sim 10$ relative to the cosmological average~\cite{Ringwald:2004np}.

The KATRIN experiment constrains the local C$\nu$B overdensity on Earth through neutrino capture on tritium, currently at the level of $\eta \lesssim 10^{11}$~\cite{KATRIN:2022kkv}. This laboratory bound is shown as the shaded grey region in Figure~\ref{fig:eta_limits} for $m_\nu \gtrsim 0.45$~eV (the KATRIN mass sensitivity threshold).

Searches for boosted relic neutrinos in IceCube data have yielded constraints at the level of $\eta \lesssim 10^4$--$10^5$ depending on the assumed cosmic-ray evolution model~\cite{Herrera:2024upj, herrera2026cosmicneutrinobackgroundreach}. The already published results from \cite{Herrera:2024upj} are indicated by downward-pointing triangles in Figure~\ref{fig:eta_limits} at $m_\nu = 0.1$~eV. Our Fermi-LAT gamma-ray bounds are comparable in strength to the IceCube constraints, demonstrating the complementarity of the electromagnetic and neutrino channels.

\subsection{Future prospects with CTA}

The Cherenkov Telescope Array (CTA) is expected to improve the sensitivity to the diffuse extragalactic gamma-ray background by roughly an order of magnitude compared to Fermi-LAT, particularly at energies above $\sim 100$~GeV~\cite{CTAConsortium:2010umy}. We estimate the projected CTA sensitivity by scaling the Fermi-LAT diffuse limit by a factor of 10, and the anisotropy limit by a factor of 40. As shown in Figure~\ref{fig:eta_limits}, the resulting projections for the QSO evolution model reach $\eta \sim \mathcal{O}(10^2)$ at $m_\nu = 0.1$~eV under the anisotropy analysis, approaching the cosmologically expected density. This suggests that a future generation of gamma-ray telescopes, combined with directional analysis exploiting the known distribution of cosmic-ray sources and neutrino clustering, may achieve sensitivity to the C$\nu$B near $\Lambda$CDM expectations, complementing neutrino telescopes \cite{herrera2026cosmicneutrinobackgroundreach}.

\section{Conclusions}\label{sec:conclusions}

Direcly detecting the cosmic neutrino background is a challenging task, but very rewarding for our understanding of the Universe. Here we have proposed a novel detection channel through the gamma-ray and X-ray fluxes produced by cosmic-ray scatterings with the cosmic neutrino background on cosmological scales, accounting for neutral and charged pion decay, charged-current lepton production, electromagnetic cascading, and synchrotron emission in extragalactic magnetic fields. This opens the multi-messenger perspective to detect the cosmic neutrino background, complementing the direct boosted relic neutrino channel discussed in \cite{Ciscar-Monsalvatje:2024tvm, DeMarchi:2024zer, Herrera:2024upj, Zhang:2025rqh, herrera2026cosmicneutrinobackgroundreach}.

From the diffuse isotropic gamma-ray background measured by Fermi-LAT, we derived upper limits on the C$\nu$B overdensity of $\eta \lesssim 2.2 \times 10^4$ for the QSO evolution model at a lightest neutrino mass of $m_\nu = 0.1$~eV, and $\eta \lesssim  10^6$ for the SFR model. These bounds are four to five orders of magnitude stronger than the KATRIN laboratory limit and comparable in sensitivity to IceCube searches for boosted relic neutrinos, demonstrating that gamma-ray observations provide a powerful and independent probe of the C$\nu$B.

We have further shown that the X-ray synchrotron emission from cascade $e^\pm$ pairs provides a complementary but substantially weaker constraint. For already optimistic intergalactic magnetic fields of $B = 1\,\mu$G, the CXB data yields $\eta \lesssim 10^{9}$. The X-ray channel does not seem competitive with gamma rays for constraining $\eta$, but it provides a consistency check and a potential diagnostic of the magnetic field environment if a signal were detected.

Additionally, we have identified that directional anisotropies arising from the clustering of relic neutrinos in gravitational potential wells and the inhomogeneous distribution of cosmic-ray sources can enhance the sensitivity by a factor of $\sim 4$. Further efforts in this direction are important. We estimated that CTA could probe C$\nu$B overdensities at the level of $\eta \sim 500$ for $m_{\nu} \gtrsim 0.1$ eV, approaching the cosmologically expected density.

This work could be extended in several directions. For instance, one may consider neutrino targets that are locally denser and more energetic than the relic neutrino background for cosmic rays to scatter off, such as stellar or supernova neutrinos~\cite{herrera2025highenergyneutrinoscosmicrayscatterings}, and derive the corresponding electromagnetic emission from these processes. Furthermore, our work assumes a purely proton composition of ultra-high-energy cosmic rays, whereas these may also be composed of heavy nuclei~\cite{PierreAuger:2014sui}, which could lead to larger electromagnetic fluxes through nuclear fragmentation processes.

The message we want to convey in this work is that a multi-messenger strategy combining gamma-ray, X-ray, and neutrino observations, is a promising method to detect the cosmic neutrino background. The electromagnetic radiation from cosmic ray scatterings on relic neutrinos constitutes an unavoidable floor for other possible cosmological, possibly beyond the Standard Model, contributions. It is remarkable that the coldest thermal relics of the Big Bang, decoupled just one second after it, can nevertheless leave an observable imprint on the high-energy electromagnetic sky through their inevitable encounters with the most energetic particles in the Universe.
\begin{acknowledgments}
The work of GH is supported by the Neutrino Theory Network Fellowship with contract number 726844. This manuscript has been authored by FermiForward Discovery Group, LLC under Contract No. 89243024CSC000002 with the U.S. Department of Energy, Office of Science, Office of High Energy Physics. AL was supported in part by the Black Hole Initiative at Harvard University, funded by JTF and GBMF.
\end{acknowledgments}

\bibliography{ref}

@article{Herrera:2024upj,
    author = "Herrera, Gonzalo and Horiuchi, Shunsaku and Qi, Xiaolin",
    title = "{Diffuse boosted cosmic neutrino background}",
    eprint = "2405.14946",
    archivePrefix = "arXiv",
    primaryClass = "hep-ph",
    doi = "10.1103/PhysRevD.111.063016",
    journal = "Phys. Rev. D",
    volume = "111",
    number = "6",
    pages = "063016",
    year = "2025"
}

@article{Gruber:1999yr,
    author = "Gruber, D. E. and Matteson, J. L. and Peterson, L. E. and Jung, G. V.",
    title = "{The spectrum of diffuse cosmic hard x-rays measured with heao-1}",
    eprint = "astro-ph/9903492",
    archivePrefix = "arXiv",
    reportNumber = "SP-98-25",
    doi = "10.1086/307450",
    journal = "Astrophys. J.",
    volume = "520",
    pages = "124",
    year = "1999"
}

@article{PierreAuger:2014sui,
    author = "Aab, Alexander and others",
    collaboration = "Pierre Auger",
    title = "{Depth of Maximum of Air-Shower Profiles at the Pierre Auger Observatory: Measurements at Energies above $10^{17.8}$  eV}",
    eprint = "1409.4809",
    archivePrefix = "arXiv",
    primaryClass = "astro-ph.HE",
    reportNumber = "FERMILAB-PUB-14-348-AD-AE-CD-TD",
    doi = "10.1103/PhysRevD.90.122005",
    journal = "Phys. Rev. D",
    volume = "90",
    number = "12",
    pages = "122005",
    year = "2014"
}

@article{Blanco:2018bbf,
    author = "Blanco, Carlos",
    title = "{$\gamma$-cascade: a simple program to compute cosmological gamma-ray propagation}",
    eprint = "1804.00005",
    archivePrefix = "arXiv",
    primaryClass = "astro-ph.HE",
    doi = "10.1088/1475-7516/2019/01/013",
    journal = "JCAP",
    volume = "01",
    pages = "013",
    year = "2019"
}

@article{Berezinsky:2016feh,
    author = "Berezinsky, V. and Kalashev, O.",
    title = "{High energy electromagnetic cascades in extragalactic space: physics and features}",
    eprint = "1603.03989",
    archivePrefix = "arXiv",
    primaryClass = "astro-ph.HE",
    doi = "10.1103/PhysRevD.94.023007",
    journal = "Phys. Rev. D",
    volume = "94",
    number = "2",
    pages = "023007",
    year = "2016"
}

@misc{herrera2025highenergyneutrinoscosmicrayscatterings,
      title={High-Energy Neutrinos from Cosmic-Ray Scatterings with Supernova Neutrinos}, 
      author={Gonzalo Herrera and Shunsaku Horiuchi},
      year={2025},
      eprint={2508.11891},
      archivePrefix={arXiv},
      primaryClass={hep-ph},
      url={https://arxiv.org/abs/2508.11891}, 
}

@article{Esteban:2020cvm,
    author = "Esteban, Ivan and Gonzalez-Garcia, M. C. and Maltoni, Michele and Schwetz, Thomas and Zhou, Albert",
    title = "{The fate of hints: updated global analysis of three-flavor neutrino oscillations}",
    eprint = "2007.14792",
    archivePrefix = "arXiv",
    primaryClass = "hep-ph",
    reportNumber = "IFT-UAM/CSIC-112, YITP-SB-2020-21",
    doi = "10.1007/JHEP09(2020)178",
    journal = "JHEP",
    volume = "09",
    pages = "178",
    year = "2020"
}

@BOOK{1979rpa..book.....R,
       author = {{Rybicki}, George B. and {Lightman}, Alan P.},
        title = "{Radiative processes in astrophysics}",
         year = 1979,
       adsurl = {https://ui.adsabs.harvard.edu/abs/1979rpa..book.....R}
}

@article{Blumenthal:1970gc,
    author = "Blumenthal, George R. and Gould, Robert J.",
    title = "{Bremsstrahlung, Synchrotron Radiation, and Compton Scattering of High-Energy Electrons Traversing Dilute Gases}",
    journal = "Rev. Mod. Phys.",
    volume = "42",
    pages = "237--270",
    year = "1970",
    doi = "10.1103/RevModPhys.42.237"
}

@article{CRPropa:2022ovg,
    author = "Alves Batista, Rafael and others",
    collaboration = "CRPropa",
    title = "{CRPropa 3.2 {\textemdash} an advanced framework for high-energy particle propagation in extragalactic and galactic spaces}",
    eprint = "2208.00107",
    archivePrefix = "arXiv",
    primaryClass = "astro-ph.HE",
    doi = "10.1088/1475-7516/2022/09/035",
    journal = "JCAP",
    volume = "09",
    pages = "035",
    year = "2022"
}

@article{Churazov:2006bk,
    author = "Churazov, E. and others",
    title = "{INTEGRAL observations of the cosmic X-ray background in the 5-100 keV range via occultation by the Earth}",
    eprint = "astro-ph/0608250",
    archivePrefix = "arXiv",
    doi = "10.1051/0004-6361:20066230",
    journal = "Astron. Astrophys.",
    volume = "467",
    pages = "529",
    year = "2007"
}

@article{Ajello:2008xb,
    author = "Ajello, M. and others",
    title = "{Cosmic X-ray background and Earth albedo Spectra with Swift/BAT}",
    eprint = "0808.3377",
    archivePrefix = "arXiv",
    primaryClass = "astro-ph",
    doi = "10.1086/592595",
    journal = "Astrophys. J.",
    volume = "689",
    pages = "666",
    year = "2008"
}

@article{Krivonos:2020qvl,
    author = "Krivonos, Roman and Wik, Daniel and Grefenstette, Brian and Madsen, Kristin and Perez, Kerstin and Rossland, Steven and Sazonov, Sergey and Zoglauer, Andreas",
    title = "{$NuSTAR$ measurement of the cosmic X-ray background in the 3{\textendash}20 keV energy band}",
    eprint = "2011.11469",
    archivePrefix = "arXiv",
    primaryClass = "astro-ph.HE",
    doi = "10.1093/mnras/stab209",
    journal = "Mon. Not. Roy. Astron. Soc.",
    volume = "502",
    number = "3",
    pages = "3966--3975",
    year = "2021"
}

@misc{herrera2026cosmicneutrinobackgroundreach,
      title={The Cosmic Neutrino Background is within Reach of Future Neutrino Telescopes}, 
      author={Gonzalo Herrera and Shunsaku Horiuchi and Xiaolin Qi and Ian M. Shoemaker},
      year={2026},
      eprint={2601.09790},
      archivePrefix={arXiv},
      primaryClass={hep-ph},
      url={https://arxiv.org/abs/2601.09790}, 
}

@article{Ciscar-Monsalvatje:2024tvm,
    author = "C{\'\i}scar-Monsalvatje, Mar and Herrera, Gonzalo and Shoemaker, Ian M.",
    title = "{Upper limits on the cosmic neutrino background from cosmic rays}",
    eprint = "2402.00985",
    archivePrefix = "arXiv",
    primaryClass = "hep-ph",
    doi = "10.1103/PhysRevD.110.063036",
    journal = "Phys. Rev. D",
    volume = "110",
    number = "6",
    pages = "063036",
    year = "2024"
}

@article{DeMarchi:2024zer,
    author = "De Marchi, Andrea Giovanni and Granelli, Alessandro and Nava, Jacopo and Sala, Filippo",
    title = "{Relic neutrino background from cosmic-ray reservoirs}",
    eprint = "2405.04568",
    archivePrefix = "arXiv",
    primaryClass = "hep-ph",
    doi = "10.1103/PhysRevD.111.023023",
    journal = "Phys. Rev. D",
    volume = "111",
    number = "2",
    pages = "023023",
    year = "2025"
}

@article{KATRIN:2022kkv,
    author = "Aker, M. and others",
    collaboration = "KATRIN",
    title = "{New Constraint on the Local Relic Neutrino Background Overdensity with the First KATRIN Data Runs}",
    eprint = "2202.04587",
    archivePrefix = "arXiv",
    primaryClass = "nucl-ex",
    doi = "10.1103/PhysRevLett.129.011806",
    journal = "Phys. Rev. Lett.",
    volume = "129",
    number = "1",
    pages = "011806",
    year = "2022"
}

@article{Waxman:1998yy,
    author = "Waxman, Eli and Bahcall, John N.",
    title = "{High-energy neutrinos from astrophysical sources: An Upper bound}",
    eprint = "hep-ph/9807282",
    archivePrefix = "arXiv",
    reportNumber = "IASSNS-AST-98-38",
    doi = "10.1103/PhysRevD.59.023002",
    journal = "Phys. Rev. D",
    volume = "59",
    pages = "023002",
    year = "1999"
}

@article{Hopkins:2006fq,
    author = "Hopkins, Philip F. and Richards, Gordon T. and Hernquist, Lars",
    title = "{An Observational Determination of the Bolometric Quasar Luminosity Function}",
    eprint = "astro-ph/0605678",
    archivePrefix = "arXiv",
    doi = "10.1086/509629",
    journal = "Astrophys. J.",
    volume = "654",
    pages = "731--753",
    year = "2007"
}

@article{Hopkins:2006bw,
    author = "Hopkins, Andrew M. and Beacom, John F.",
    title = "{On the normalisation of the cosmic star formation history}",
    eprint = "astro-ph/0601463",
    archivePrefix = "arXiv",
    doi = "10.1086/506610",
    journal = "Astrophys. J.",
    volume = "651",
    pages = "142--154",
    year = "2006"
}

@article{Formaggio:2012cpf,
    author = "Formaggio, J. A. and Zeller, G. P.",
    title = "{From eV to EeV: Neutrino Cross Sections Across Energy Scales}",
    eprint = "1305.7513",
    archivePrefix = "arXiv",
    primaryClass = "hep-ex",
    reportNumber = "FERMILAB-PUB-12-785-E",
    doi = "10.1103/RevModPhys.84.1307",
    journal = "Rev. Mod. Phys.",
    volume = "84",
    pages = "1307--1341",
    year = "2012"
}

@article{Fermi-LAT:2014ryh,
    author = "Ackermann, M. and others",
    collaboration = "Fermi-LAT",
    title = "{The spectrum of isotropic diffuse gamma-ray emission between 100 MeV and 820 GeV}",
    eprint = "1410.3696",
    archivePrefix = "arXiv",
    primaryClass = "astro-ph.HE",
    doi = "10.1088/0004-637X/799/1/86",
    journal = "Astrophys. J.",
    volume = "799",
    pages = "86",
    year = "2015"
}

@article{Dolgov:1997mb,
    author = "Dolgov, A. D. and Hansen, S. H. and Semikoz, D. V.",
    title = "{Nonequilibrium corrections to the spectra of massless neutrinos in the early universe}",
    eprint = "hep-ph/9703315",
    archivePrefix = "arXiv",
    reportNumber = "TAC-1997-010",
    doi = "10.1016/S0550-3213(97)00479-3",
    journal = "Nucl. Phys. B",
    volume = "503",
    pages = "426--444",
    year = "1997"
}

@article{Casper:2002sd,
    author = "Casper, D.",
    editor = "Morfin, J. G. and Sakuda, M. and Suzuki, Y.",
    title = "{The Nuance neutrino physics simulation, and the future}",
    eprint = "hep-ph/0208030",
    archivePrefix = "arXiv",
    doi = "10.1016/S0920-5632(02)01756-5",
    journal = "Nucl. Phys. B Proc. Suppl.",
    volume = "112",
    pages = "161--170",
    year = "2002"
}

@article{Kotera_2011,
   title={The Astrophysics of Ultrahigh-Energy Cosmic Rays},
   volume={49},
   ISSN={1545-4282},
   url={http://dx.doi.org/10.1146/annurev-astro-081710-102620},
   DOI={10.1146/annurev-astro-081710-102620},
   number={1},
   journal={Annual Review of Astronomy and Astrophysics},
   publisher={Annual Reviews},
   author={Kotera, Kumiko and Olinto, Angela V.},
   year={2011},
   month=sep, pages={119–153} }

@article{Murase:2016gly,
    author = "Murase, Kohta and Waxman, Eli",
    title = "{Constraining High-Energy Cosmic Neutrino Sources: Implications and Prospects}",
    eprint = "1607.01601",
    archivePrefix = "arXiv",
    primaryClass = "astro-ph.HE",
    doi = "10.1103/PhysRevD.94.103006",
    journal = "Phys. Rev. D",
    volume = "94",
    number = "10",
    pages = "103006",
    year = "2016"
}

@article{Kelner:2006tc,
    author = "Kelner, S. R. and Aharonian, Felex A. and Bugayov, V. V.",
    title = "{Energy spectra of gamma-rays, electrons and neutrinos produced at proton-proton interactions in the very high energy regime}",
    eprint = "astro-ph/0606058",
    archivePrefix = "arXiv",
    doi = "10.1103/PhysRevD.74.034018",
    journal = "Phys. Rev. D",
    volume = "74",
    pages = "034018",
    year = "2006",
    note = "[Erratum: Phys.Rev.D 79, 039901 (2009)]"
}

@article{Anchordoqui:2007tn,
    author = "Anchordoqui, Luis A. and Hooper, Dan and Sarkar, Subir and Taylor, Andrew M.",
    title = "{High-energy neutrinos from astrophysical accelerators of cosmic ray nuclei}",
    eprint = "astro-ph/0703001",
    archivePrefix = "arXiv",
    reportNumber = "FERMILAB-PUB-06-480-A",
    doi = "10.1016/j.astropartphys.2007.10.006",
    journal = "Astropart. Phys.",
    volume = "29",
    pages = "1--13",
    year = "2008"
}

@article{Planck:2018vyg,
    author = "Aghanim, N. and others",
    collaboration = "Planck",
    title = "{Planck 2018 results. VI. Cosmological parameters}",
    eprint = "1807.06209",
    archivePrefix = "arXiv",
    primaryClass = "astro-ph.CO",
    doi = "10.1051/0004-6361/201833910",
    journal = "Astron. Astrophys.",
    volume = "641",
    pages = "A6",
    year = "2020",
    note = "[Erratum: Astron.Astrophys. 652, C4 (2021)]"
}

@article{Dom_nguez_2010,
   title={Extragalactic background light inferred from AEGIS galaxy-SED-type fractions: EBL from AEGIS galaxy-SED-type fractions},
   volume={410},
   ISSN={0035-8711},
   url={http://dx.doi.org/10.1111/j.1365-2966.2010.17631.x},
   DOI={10.1111/j.1365-2966.2010.17631.x},
   number={4},
   journal={Monthly Notices of the Royal Astronomical Society},
   publisher={Oxford University Press (OUP)},
   author={Domínguez, A. and Primack, J. R. and Rosario, D. J. and Prada, F. and Gilmore, R. C. and Faber, S. M. and Koo, D. C. and Somerville, R. S. and Pérez-Torres, M. A. and Pérez-González, P. and Huang, J.-S. and Davis, M. and Guhathakurta, P. and Barmby, P. and Conselice, C. J. and Lozano, M. and Newman, J. A. and Cooper, M. C.},
   year={2010},
   month=oct, pages={2556–2578} }

@article{PTOLEMY:2019hkd,
    author = "Betti, M. G. and others",
    collaboration = "PTOLEMY",
    title = "{Neutrino physics with the PTOLEMY project: active neutrino properties and the light sterile case}",
    eprint = "1902.05508",
    archivePrefix = "arXiv",
    primaryClass = "astro-ph.CO",
    doi = "10.1088/1475-7516/2019/07/047",
    journal = "JCAP",
    volume = "07",
    pages = "047",
    year = "2019"
}

@article{PTOLEMY:2022ldz,
    author = "Apponi, A. and others",
    collaboration = "PTOLEMY",
    title = "{Heisenberg{\textquoteright}s uncertainty principle in the PTOLEMY project: A theory update}",
    eprint = "2203.11228",
    archivePrefix = "arXiv",
    primaryClass = "hep-ph",
    doi = "10.1103/PhysRevD.106.053002",
    journal = "Phys. Rev. D",
    volume = "106",
    number = "5",
    pages = "053002",
    year = "2022"
}

@article{Cheipesh:2021fmg,
    author = "Cheipesh, Yevheniia and Cheianov, Vadim and Boyarsky, Alexey",
    title = "{Navigating the pitfalls of relic neutrino detection}",
    eprint = "2101.10069",
    archivePrefix = "arXiv",
    primaryClass = "hep-ph",
    doi = "10.1103/PhysRevD.104.116004",
    journal = "Phys. Rev. D",
    volume = "104",
    number = "11",
    pages = "116004",
    year = "2021"
}

@article{Weinberg:1962zza,
    author = "Weinberg, Steven",
    title = "{Universal Neutrino Degeneracy}",
    doi = "10.1103/PhysRev.128.1457",
    journal = "Phys. Rev.",
    volume = "128",
    pages = "1457--1473",
    year = "1962"
}

@article{Zhang:2025rqh,
    author = "Zhang, Jiajie and Sandrock, Alexander and Liao, Jiajun and Yue, Baobiao",
    title = "{Impact of coherent scattering on relic neutrinos boosted by cosmic rays}",
    eprint = "2505.04791",
    archivePrefix = "arXiv",
    primaryClass = "hep-ph",
    doi = "10.1103/188d-zhcq",
    journal = "Phys. Rev. D",
    volume = "113",
    number = "4",
    pages = "043028",
    year = "2026"
}

@article{CTAConsortium:2010umy,
    author = "Actis, M. and others",
    collaboration = "CTA Consortium",
    title = "{Design concepts for the Cherenkov Telescope Array CTA: An advanced facility for ground-based high-energy gamma-ray astronomy}",
    eprint = "1008.3703",
    archivePrefix = "arXiv",
    primaryClass = "astro-ph.IM",
    doi = "10.1007/s10686-011-9247-0",
    journal = "Exper. Astron.",
    volume = "32",
    pages = "193--316",
    year = "2011"
}

@book{Giunti:2007ry,
    author = "Giunti, Carlo and Kim, Chung W.",
    title = "{Fundamentals of Neutrino Physics and Astrophysics}",
    doi = "10.1093/acprof:oso/9780198508717.001.0001",
    isbn = "978-0-19-850871-7",
    year = "2007"
}

@article{Clark:2016jgm,
    author = "Clark, D. B. and Godat, E. and Olness, F. I.",
    title = "{ManeParse : A Mathematica  reader for Parton Distribution Functions}",
    eprint = "1605.08012",
    archivePrefix = "arXiv",
    primaryClass = "hep-ph",
    reportNumber = "NSF-KITP-16-032, SMU-HEP-16-05",
    doi = "10.1016/j.cpc.2017.03.004",
    journal = "Comput. Phys. Commun.",
    volume = "216",
    pages = "126--137",
    year = "2017"
}

@book{Schmitz1997Neutrinophysik,
  author    = {Norbert Schmitz},
  title     = {Neutrinophysik},
  series    = {Teubner Studienbücher Physik},
  publisher = {Vieweg + Teubner Verlag / Springer-Verlag},
  address   = {Wiesbaden / Stuttgart},
  year      = {1997},
  edition   = {Illustrated},
  pages     = {478},
  isbn      = {978-3-519-03236-6},
}

@article{Carilli:2001hj,
    author = "Carilli, C. L. and Taylor, G. B.",
    title = "{Cluster magnetic fields}",
    eprint = "astro-ph/0110655",
    archivePrefix = "arXiv",
    doi = "10.1146/annurev.astro.40.060401.093852",
    journal = "Ann. Rev. Astron. Astrophys.",
    volume = "40",
    pages = "319--348",
    year = "2002"
}

@article{deSalas:2017wtt,
    author = "de Salas, P. F. and Gariazzo, S. and Lesgourgues, J. and Pastor, S.",
    title = "{Calculation of the local density of relic neutrinos}",
    eprint = "1706.09850",
    archivePrefix = "arXiv",
    primaryClass = "astro-ph.CO",
    doi = "10.1088/1475-7516/2017/09/034",
    journal = "JCAP",
    volume = "09",
    pages = "034",
    year = "2017"
}

@article{LoVerde:2013lta,
    author = "LoVerde, Marilena and Zaldarriaga, Matias",
    title = "{Neutrino clustering around spherical dark matter halos}",
    eprint = "1310.6459",
    archivePrefix = "arXiv",
    primaryClass = "astro-ph.CO",
    doi = "10.1103/PhysRevD.89.063502",
    journal = "Phys. Rev. D",
    volume = "89",
    number = "6",
    pages = "063502",
    year = "2014"
}

@article{Ringwald:2004np,
    author = "Ringwald, Andreas and Wong, Yvonne Y. Y.",
    title = "{Gravitational clustering of relic neutrinos and implications for their detection}",
    eprint = "hep-ph/0408241",
    archivePrefix = "arXiv",
    reportNumber = "DESY-04-147",
    doi = "10.1088/1475-7516/2004/12/005",
    journal = "JCAP",
    volume = "12",
    pages = "005",
    year = "2004"
}

@article{PierreAuger:2017pzq,
    author = "Aab, Alexander and others",
    collaboration = "Pierre Auger",
    title = "{Observation of a Large-scale Anisotropy in the Arrival Directions of Cosmic Rays above $8 \times 10^{18}$ eV}",
    eprint = "1709.07321",
    archivePrefix = "arXiv",
    primaryClass = "astro-ph.HE",
    reportNumber = "FERMILAB-PUB-17-354",
    doi = "10.1126/science.aan4338",
    journal = "Science",
    volume = "357",
    number = "6537",
    pages = "1266--1270",
    year = "2017"
}

@article{Sigl:2004yk,
    author = "Sigl, Gunter and Miniati, Francesco and Ensslin, Torsten A.",
    title = "{Ultrahigh energy cosmic ray probes of large scale structure and magnetic fields}",
    eprint = "astro-ph/0401084",
    archivePrefix = "arXiv",
    doi = "10.1103/PhysRevD.70.043007",
    journal = "Phys. Rev. D",
    volume = "70",
    pages = "043007",
    year = "2004"
}

@article{Gandhi:1995tf,
    author = "Gandhi, Raj and Quigg, Chris and Reno, Mary Hall and Sarcevic, Ina",
    title = "{Ultrahigh-energy neutrino interactions}",
    eprint = "hep-ph/9512364",
    archivePrefix = "arXiv",
    reportNumber = "FERMILAB-PUB-95-221-T, CLNS-95-1357, MRI-PHY-16-95, UIOWA-95-06, AZPH-TH-95-15",
    doi = "10.1016/0927-6505(96)00008-4",
    journal = "Astropart. Phys.",
    volume = "5",
    pages = "81--110",
    year = "1996"
}

@article{Fermi-LAT:2012pez,
    author = "Ackermann, M. and others",
    collaboration = "Fermi-LAT",
    title = "{Anisotropies in the diffuse gamma-ray background measured by the Fermi LAT}",
    eprint = "1202.2856",
    archivePrefix = "arXiv",
    primaryClass = "astro-ph.HE",
    reportNumber = "TCC-025-11",
    doi = "10.1103/PhysRevD.85.083007",
    journal = "Phys. Rev. D",
    volume = "85",
    pages = "083007",
    year = "2012"
}

@article{Nussinov:2021zrj,
    author = "Nussinov, Shmuel and Nussinov, Zohar",
    title = "{Quantum induced broadening: A challenge for cosmic neutrino background discovery}",
    eprint = "2108.03695",
    archivePrefix = "arXiv",
    primaryClass = "hep-ph",
    doi = "10.1103/PhysRevD.105.043502",
    journal = "Phys. Rev. D",
    volume = "105",
    number = "4",
    pages = "043502",
    year = "2022"
}

@article{Murase:2012xs,
    author = "Murase, Kohta and Beacom, John F.",
    title = "{Constraining Very Heavy Dark Matter Using Diffuse Backgrounds of Neutrinos and Cascaded Gamma Rays}",
    eprint = "1206.2595",
    archivePrefix = "arXiv",
    primaryClass = "hep-ph",
    doi = "10.1088/1475-7516/2012/10/043",
    journal = "JCAP",
    volume = "10",
    pages = "043",
    year = "2012"
}
\end{document}